\documentclass[aps,pre,preprint]{revtex4}

\usepackage{graphicx}
\usepackage[breaklinks=true,urlcolor=blue,bookmarks=true]{hyperref}
\usepackage{docs}

\begin{document}
\title{Dielectric Enhancement from Non-Insulating Particles with Ideally 
Polarized Interfaces and Zero $\zeta$-Potential I: Exact Solution}
\author{ Jiang Qian$^1$ and Pabitra N Sen$^2$ }
\affiliation{ $^1$  4097 Silsby Road, University Heights, OH 44118 \\$^2$ 35 Hazel Road, Berkeley, Ca 94705  }
\date{\today}
\begin{abstract}
	We solve exactly the dielectric response of a \emph{non-insulating} 
sphere of radius $a$ suspended in symmetric, univalent electrolyte solution, 
with \emph{ideally-polarizable} interface but \emph{without} significant 
$\zeta$-potential. We then use this solution to derive the dielectric response 
of a dilute random suspension of such spheres, with volume fraction $f\ll1$, 
within the Maxwell-Garnett Effective Medium Approximation. Surprisingly, we 
discover a huge dielectric enhancement in this bare essential model of 
dielectric responses of solids in electrolyte solution: at low frequency 
$\omega\tau_D\ll(\lambda/a)/(\sigma_w/\sigma_s+1/2)$, the real part of the 
effective dielectric constant of the mixture is $1-(3f/2)+(9f/4)(a/\lambda)$.  
Here $\sigma_{w/s}$ is the conductivity of the electrolyte solution/solids, 
$\lambda$ is the Debye screening length in the solution, $\tau_D=\lambda^2/D$ 
is the standard time scale of diffusion and $D$ is the ion
diffusion coefficient. As $\lambda$ is of the order nm even for dilute 
electrolyte solution, even for sub-mm spheres and low volume fraction $f=0.05$ 
the huge geometric factor $a/\lambda$ implies an over $10^4$-fold enhancement.  
Furthermore, we show that this enhancement produces a significant low frequency 
($\omega\tau_D\ll1$) phase shift 
$\tan\theta=\mathrm{Re}~\epsilon(\omega)/\mathrm{Im}~\epsilon(\omega)$ in a 
simple impedance measurement of the mixture, which is usually negligible in 
pure electrolyte solution. The phase shift has a \emph{scale-invariant} maximum 
$\tan\theta_{\mathrm{max}}=(9/4)f/(2\sigma_w/\sigma_s+1)$ at 
$\omega_{\mathrm{max}}=(2D/\lambda a)/(2\sigma_w/\sigma_s+1)$. We provide a 
physical picture of the enhancement from an accumulation of charges in a thin 
\underline{E}xternally \underline{I}nduced \underline{D}ouble \underline{L}ayer 
(EIDL) due to the blocking boundary condition on interfaces. This mechanism is 
distinct from the traditional dielectric enhancement in \emph{insulating} 
particles due to large intrinsic $\zeta$-potentials and surface charges, which 
predicts a different scaling for maximum phase shift frequency 
$\omega_{\mathrm{max}}=D/a^2$. Our model is also more transparent than that of 
Wong~(\emph{Geophysics} 44, 1245), which invokes different types of ions with 
or without Faradaic currents that obscure the physics behind his results.  
Finally we discuss data from geological samples containing sulfides and recent 
experiments on coke freeze that comport well with our predictions.
\end{abstract}
\maketitle
\section{Introduction}\label{sec:intro}
In this paper we show, using an \emph{exactly solvable model}, that a dilute 
suspension in electrolyte solution of non-insulating solid spheres with ideally 
polarized interfaces, across which no charge, electronic or ionic, is 
transferred between the solution and the solid, can give rise to huge 
dielectric constants ($10^{5-6}$ or more) at low frequencies (KHz, far below 
plasma frequencies).  This is remarkable considering that the electrolyte 
static permittivity is only about 80 while particle permittivity (due to core 
electrons and lattice) is about 1-5 at such frequencies.  The only similar 
prior result we are aware of is by Wong \cite{Wong}, who used a fixed (zero)  
potential interface, which generally means an ideally \emph{non}-polarizable 
interface \cite{Bard} held at a constant potential by the redox reactions. But 
more important, Wong's analysis involved a medley of different kinds of ions,
some with Faradaic  (redox) reactions at the interface and other without. This  
made the physics  obscure and source of the enhancement in his model is not 
clear.  Here  we strip down to essential physics and use an exactly solved 
model, derived from first principle, for \emph{ideally polarizable interface} 
to show that dielectric enhancements are natural outcome in non-insulating 
particles in electrolyte solution, explicable without resorting to complex 
surface chemistry.

The enhancement here are produced on  interfaces with zero $\zeta$-potential.  
The dielectric enhancements with  \emph{insulating} clay-like charged particles 
with finite $\zeta$-potentials suspended in electrolytes  are well known 
\cite{Dukhin, Fixman, Chew, Hinch, Others}, and are due to the Guoy-Chapman 
layer of the counter ions.  Briefly,    dry clay   has a dielectric constant of 
about 5 and that of water is about 80 but  the real part  the dielectric in a 
clay - brine mixture   can be as large as   $\sim 10^4$ at low frequencies 
\cite{Dukhin, Fixman, Chew, Hinch, Others, Schwan_enhancement,Clay,Lyklema}.  
Standard Maxwell-Wagner / Maxwell-Garnett 
(\cite{Maxwell,MaxwellGarnett,Landauer, Batchelor}) analysis for a two-phase 
material is unable to explain such enhancements, see the Appendix I. Hinch 
\emph{et. al.}~\cite{Hinch} have nicely summarized the criticisms of previous 
ad hoc models~\cite{Schwan_enhancement} that ``explained" the enhancement by  
endowing  the particles with capacitative layers or layers with complex 
conductivity that could not be calculated from the first principles. 

This clay-effect for insulating particles disappears when the $\zeta$-potential 
is zero.  Here we show that even  with zero $\zeta$-potential, non-insulating 
materials are capable of showing an dielectric enhancement, due to an 
\underline{E}lectrically \underline{I}nduced \underline{D}ouble 
\underline{L}ayer (EIDL) produced by the low-frequency charge build-up from the 
blocking boundaries.

Analysis of electro-capillarity, electrophoresis and electro viscous forces on  
ideally polarizable metal drops  require computing the polarizability of 
metallic spheres as we do--see the  book by Levich \cite{Levich} and references 
therein, and more recent papers \cite{Bazant,Ohshima, Murtsovkin}.The term EIDL 
has been used by Bazant and his co-workers \cite{Bazant} in their study of 
electrokinetic effects.  However these works do not consider the dielectric 
enhancement. Also, the previous work restrict themselves to the case of 
infinite particle conductivity, \emph{i.e.},  an ideally polarizable surface 
that is an equipotential, with infinite conductivity for the metal particle.  
In our case we tackle solids of all conductivities, and indeed, we show below 
that the characteristic frequency for the enhancement depends on the ratio of 
the electrical conductivity of the particle and that of the electrolyte. We 
have used Levich's insight \cite{Levich} to estimate the effect of 
electrokinetic flow on surface physics.  In Ref.  (\cite{Jiang}) we show that 
the electrophoresis does not affect the physics in the case of negligible 
$\zeta$-potential. By contrast, for the finite zeta potential in clay like 
particles, the electrokinetic effect can be large--see, for example, Hinch 
\emph{et. al.}~\cite{Hinch} or Fixman\cite{Fixman}. We will thus not consider 
the electrokinetic effects here.

Electrochemistry touches many area of science and technology, ranging from 
biology, energy conversion, hydrolysis, batteries, corrosion in ships, 
bio-implants, pipeline corrosion,  to geophysical explorations for minerals and 
hydrocarbon.  One of the paradigm of interface is an ideally polarizable 
interface, the other being an ideally non-polarizable interface. The gap in the 
literature in understanding the enhancement in dielectric for the ideally 
polarizable non-insulating materials is glaring, because the interfacial  
processes at the solid-fluid interface have fundamental  implications in the 
basic science of electrochemistry.  

Noble metals, a common type of ideally-polarizable materials, are often used as 
contrast agent and markers in medicine.  The optical properties of gold 
colloids are often exploited in the laboratory for markers in cancer cells. The 
possibility of using low frequency electrical property predicted here, where 
the depth of penetration is large, could be useful in clinical settings   not 
unlike Electroencephalography imaging or impedance tomography\cite{Boon} as in  
geophysical prospecting\cite{UNC Patent, Anderson, Nelson, Seigel, Wong}. 

In geophysics deep-look using electromagnetic methods is of paramount 
interest\cite{UNC Patent, Anderson, Nelson, Seigel, Wong}. First, the so-called 
Induced Polarization (IP)  or high dielectric constant due to metallic deposits 
are routinely used in mineral exploration for gold  and copper. See, for 
example, references in Seigel \cite{Seigel} and Wong \cite{Wong}.  Seigel 
\emph{et. al.}~\cite{Seigel, Nelson}  trace the early development of the 
induced polarization method, starting with field observations by Conrad 
Schlumberger in a mining region in France in 1913.  The enhanced dielectric 
constant  has been seen in numerous samples containing pyrite, chalcopyrite 
\emph{etc.}~\cite{UNC Patent, Anderson, Nelson, Seigel, Wong}, and more 
recently on graphitic materials like coke-breeze\cite{UNC Patent}.

The second application is locating kerogeneous (source rock) material in shale 
gas and shale oil exploration\cite{Anderson} via the elctromagnetic response of 
pyrite nougats that often accompany the source. In the absence of a basic 
understanding of IP, people resort to  ad hoc fitting routines like Cole-Cole 
plot \cite{Wong, Nelson}. This work shows that the IP effect can be understood 
using fundamental physics.  It may be possible to estimate the volume fraction 
as well as the size of the metallic nougats. These size and volume fraction 
parameters, in turn, contain valuable information of  the geologic processes, 
such as the degree of maturity and the reducing nature of the depositional 
environment. Such information is useful for evaluation of the reservoir 
potential.     

The third example entails using metallic grains as contrast agent in mapping 
fractures using dielectric  tomography \cite{UNC Patent}. Hydraulic fractures 
in a  gas or an oil well are induced to enhance flow and are kept opened by  
adding  sand (proppant materials) to the hydraulic fluids. Most of  the 
fractures close upon the removal of the applied hydraulic pressure. The 
fractures that remain open are the conduits of the hydrocarbon and hence 
mapping them  is highly desirable. It has been proposed that by adding   
conducting material with high IP ( ``contrast agents") to sand one can map the 
zone using various geophysical electrical methods\cite{Anderson}. 
  
In this paper we study the exactly solvable case of spherical particles.  In 
Ref.\cite{Jiang}   we show that the dielectric enhancement observed in this 
paper is universal: it holds for non-insulating particles of arbitrary shape 
suspended in an electrolyte solution with ideally polarizable interfaces with 
small $\zeta$-potential. Crucial for our results are the ideally-polarizable  
interface  that ``blocks" any dc (Faradaic) current.  Our boundary conditions 
are different from that used previously, in that, we do not use a constant 
(zero-) potential boundary condition, but instead, apply the continuity of 
potential and the continuity of the normal component of the total current that 
comprises of the sum of displacement and conduction current.  As the ions do 
not penetrate the solid, the  current outside the solid is just the 
displacement current. Inside the solid it is  a sum of the displacement current 
and the conduction current.  In the hindsight the  continuity of current seems 
obvious, but it has not been invoked previously in these problems. We show 
below these boundary conditions are crucial for the dielectric enhancement 
effect.

This paper is organized as following. In Sec.~\ref{sec:model} we derive from 
the first principle our model of solids in electrolyte solution, explaining the 
approximation involved and their justifications and paying special attention to 
the critical boundary conditions used. Next, in Sec.~\ref{sec:single} we solve 
exactly our model for a single sphere in solution, deriving the ``Debye-like'' 
form of the dielectric response. Building on this, in Sec.~\ref{sec:suspension} 
we compute the effective complex dielectric constant for a random mixture of 
spheres and electrolyte solutions from the Maxwell-Garnett theory of Effective 
Medium Approximation. We show that it exhibits a ``Debye'' form and at low 
frequency shows a huge enhancement of the real dielectric constant. Then in 
Sec.~\ref{sec:phase} we show that this enhancement renders the previously 
negligible phase shift (or loss-angle) in a simple impedance measurement 
observable. Indeed, we show that the phase shift has a maximum at the frequency 
$\omega_{\mathrm{max}}\approx 2D/(\lambda a)$ and not at $\omega_{clay} \approx 
2D/ a^2 $ as in clay-like particles with finite zeta-potential \cite{Dukhin, 
Fixman, Chew, Hinch, Others}. For most materials, the height of the phase shift 
maximum $9f/4$ is scale-invariant and determined solely by the volume fraction.  
In Sec.~\ref{sec:physics} we use an ``effective boundary conditions'', derived 
from a picture of EIDL, to link the physics behind the enhancement to the 
ideally-polarized blocking boundary conditions in Sec.~\ref{sec:model}.  
Finally, in Sec.~\ref{sec:experiment} we discuss features of our predictions 
that have been seen in geological samples containing pyrite, chalcopyrite 
etc.\cite{Wong, Nelson}, and more recently on graphitic materials like 
coke-breeze\cite{UNC Patent}.

\section{Governing Equations and Ideally Polarized Boundary 
Conditions}\label{sec:model}
Consider dielectrically uniform solid bodies immersed in an electrolyte 
solution. A uniform external field $E_0\exp(i\omega t)$ drives the charge 
dynamics of the system. To emphasize physics and simplify notations, we assume 
the electrolyte solution
contains a single species of cations and anions, with charges $\pm e$, and that 
they share the same diffusion coefficient $D$. Elsewhere~\cite{Jiang}, we show 
that our results can be easily generalized to the cases of asymmetric ions, as 
long as the fluid is charge-neutral when $E_0=0$.

The physics inside the simple, uniform dielectric solid is entirely 
characterized by a potential $\psi_S$, obeying Laplace's equation 
$\nabla^2\psi_S=0$. The physics in the electrolyte solution is characterized by 
more complex non-uniform charge dynamics. The potential in the liquid $\psi$ 
obeys Poisson equation:
\begin{equation}
\nabla ^2 \psi( \vec{r},t)= -\frac{\rho( \vec{r},t)}{\epsilon^{\prime}_w 
\epsilon_0}; \,\,\,\,\,\,\,\, \rho ( \vec{r},t)= e (N_+( \vec{r},t) - N_-( 
\vec{r},t)),
\label{eq:poisson}
\end{equation}
where $\epsilon'_w$ is the static dielectric constant for water and $N_\pm$ are 
the ion densities.

The motion of ions in the liquid is characterized by the current densities 
$\vec{j}_N^\pm$, which consist of three components: the diffusive current 
driven by ion density gradients, the conductive current driven by the electric 
field and a hydrodynamic current from ions being carried by the macroscopic 
motion of the liquid itself. Without significant $\zeta$-potential and surface 
charge on solids, the net charge density in the solution is due entirely to the 
external field. The electrokinetic flow is governed by the 
Helmholtz-Smoluchowski equation~\cite{Levich}, and is proportional to $E_0^2$.  
Elsewhere~\cite{Jiang}, we show that the ionic currents carried by such an 
electrokinetic flows are much smaller than the conductive currents, which is 
proportional to $E_0$, and can be safely ignored in the low-frequency and 
low-field linear limit discussed in this paper.

So the ionic currents $\vec{j}^N_\pm$ in the solution consist of a diffusive 
and a conductive part, related to each other by the Einstein relation. These 
currents also determine the dynamics of ion densities through number 
conservation laws:
\begin{equation}
	\vec{j}_\pm^{\,N} =-D\left(\overrightarrow{\nabla} N_\pm \pm\frac{e 
N_\pm}{k_BT} \overrightarrow{\nabla}\psi\right),\,\,\,\,
{\overrightarrow\nabla}\cdot\vec{j}_{\pm}^{\,N} ( \vec{r},t)
= -{\partial N_{\pm}( \vec{r},t) \over \partial t} \label{eq:charge_consv}
\end{equation}
Eq.~\ref{eq:poisson} and Eq.~\ref{eq:charge_consv} fully characterize the 
physics in the electrolyte solution with a set of coupled non-linear partial 
differential equations. Major further simplifications come from the assumptions 
that both the $\zeta$-potential and the driving field are small: 
$e\psi_{\zeta}\ll k_BT$, $eE_0a\ll k_BT$, where $a$ is the maximum linear 
dimension of the solids in the direction of the $\vec{E}_0$. Under these 
conditions, we have shown elsewhere~\cite{Jiang} that, when the ion densities 
are divided into a background density $N^0_\pm=N$ in the absence of the 
external drive $E_0$ and a perturbation due to the $E_0$, $N_\pm=N+n_\pm$, the 
background is nearly uniform $\vec{\nabla}N\approx0$ and the perturbations are 
small $n_\pm\ll N$. With these assumptions, the governing equations for the 
physics in the liquid linearize:
\begin{equation}
\label{eq:motion}
{\overrightarrow\nabla}\cdot\vec{j}_{\pm}=i\omega n_{\pm},\,\,\,\,
\vec{j}_\pm =-D\left(\overrightarrow{\nabla} n_\pm \pm\frac{e N}{k_BT} 
\overrightarrow{\nabla}\psi\right),\,\,\,\,
\nabla^2 \psi=-\frac{e(n_+-n_-)}{ \epsilon_0 \epsilon_w'}.
\end{equation}

To implement the ideally polarized boundary conditions (BCs), we note that the 
ions cannot penetrate the solid, and at the same time the solid is not a source 
of ions. Furthermore, electrons do not transfer across the 
interfaces~\cite{Wong,Bard}. In this case, no charge transfer between the 
solids and the electrolyte solution occurs  at the interface, the interfaces 
are called ``perfectly polarizable" or  ``ideally polarizable". The text-book 
example of an ``ideally polarizable" interface is a platinum electrode.  For 
ideally non-polarizable interface, such as silver/silver chloride system in a 
brine,  chlorine reacts with silver/silver chloride electrode, and a Faradaic 
current can freely pass (without polarization)  through  the interface. The 
difference between ideally polarized and non-polarizable surfaces has been 
explained exceedingly well by Wong \cite{Wong}. The ideally polarized boundary 
condition means the normal components of ionic currents must vanish at 
solid-liquid surfaces:
\begin{equation}
\label{eq:bc}
\widehat{u}\cdot j_\pm |_{\mathbf{\Sigma}}
=0\,\,\,\mathbf{[a]},\,\,\,\,\,
\psi_{S}=\psi\,\,\,\mathbf{[b]},\,\,\,\,\,
(\sigma _{s}+i \omega 
\epsilon_0\epsilon_{s}')\widehat{u}\cdot\overrightarrow{\nabla}\psi_S=
(i\omega  \epsilon _w'\epsilon 
_0)\,\widehat{u}\cdot\overrightarrow{\nabla}\psi\,\,\,\mathbf{[c]}
\end{equation}
Here $\mathbf{\Sigma}$ are the liquid-solid interfaces and $\hat{u}$ is the 
normal vector on $\mathbf{\Sigma}$. $\epsilon'_s,\sigma_s$ are the 
\emph{static} dielectric constant and conductivity of solid bodies. We assume 
in this paper the frequency is much below the plasma frequencies of either the 
liquid or the solids, so that the real dielectric constants and conductivities 
can be considered frequency-independent. The conditions [{\bf{b}}]\,[{\bf{c}}] 
are just the standard BCs for potentials across dielectric interfaces. In 
particular, $[\mathbf{c}]$ is derived, as usual, from the conservation of 
currents across the interfaces $\Sigma$, with the special requirement that the 
ionic currents on the liquid side are zero due to condition $[\mathbf{a}]$, so 
there are only displacement currents on the liquid side.

The symmetry between cations and anions, though not essential to our 
conclusions, does afford a further simplification of our formalism. Introduce 
the total net ionic density $n^{\mathrm{total}}=n_++n_-$ and the net density 
$n^{\mathrm{net}}=n_-+n_-$. It is easy to transform the equations of motion 
Eq.~\ref{eq:motion} and the BCs Eq.~\ref{eq:bc} to \emph{decoupled} equations 
for $n^{\mathrm{total}}$ and $n^{\mathrm{net}}$. Furthermore, because the 
equation of motions and BCs for $n^{\mathrm{total}}$ are entirely decoupled 
from the potential $\psi$, it is easy to show $n^{\mathrm{total}}(\vec{r},t)=0$ 
throughout the liquid and at all time.

Only $n^{\mathrm{net}}$ is coupled to $\psi$ and has non-trivial dynamics.  
Combining the charge conservation and Poisson's Equation in 
Eq.~\ref{eq:motion}, the equations of motion take a very simple form:
\begin{equation}\label{eq:net_motion}
	\nabla^2 n^{\mathrm{net}}=\beta^2 n^{\mathrm{net}};\,\,\,\,\,\,\,\,
	\nabla^2 \psi=-\frac{e n^{\mathrm{net}}}{ \epsilon_0 
	\epsilon_w'};\,\,\,\,\,
	\beta^2 \lambda ^2 = 1 +i \,\omega \tau_D;\,\,\,\,\,
\tau_D=  \frac{ \lambda^2}{D}=\frac{\epsilon_0\epsilon_w'}{\sigma_w}.
\end{equation}
Here we introduce the characteristic time scale of the charge dynamics in the 
liquid $\tau_D$, which typically ranges from $10^{-10}$s to $10^{-7}$s from 
concentrated to dilute electrolyte solution, and the Debye length $\lambda$, 
which ranges from $0.3$nm to $10$nm. In this paper, we always consider low 
frequency $\omega\tau_D\ll1$, so $\beta\approx1/\lambda$. Finally, the boundary 
conditions can also be significantly simplified:
\begin{equation}\label{eq:net_BC}
	\hat{u}\cdot\vec{\nabla}n^{\mathrm{net}}+
	\frac{1}{\lambda^2}\frac{\epsilon_0\epsilon'_w}{e}
	\hat{u}\cdot\vec{\nabla}\psi=0.\,\,\mathbf{[a]},\,\,\,\,\,\,
\psi_{S}=\psi\,\,\mathbf{[b]},\,\,\,\,\,\,
\widehat{u}\cdot\overrightarrow{\nabla}\psi_S=
c\,\widehat{u}\cdot\overrightarrow{\nabla}\psi\,\,\mathbf{[c]}.
\end{equation}
Here we introduced constant $c$ to simplify notations and used the following 
form of $\lambda$, derivable from the Einstein relations:
\begin{equation}\label{eq:constants}
c=\frac{i\omega  \epsilon _w'\epsilon _0}
{\sigma _{s}+i \omega \epsilon_0\epsilon_{s}'},\,\,\,\,\,\,\,\,\,
\lambda^2 = \frac{k_BT\,\epsilon_0 \epsilon_w'}{2 N e^2}.
\end{equation}

\section{Dielectric Response of a Single Sphere in an Electrolyte 
Solution}\label{sec:single}
For a single solid sphere of radius $a$ in an electrolyte liquid, the solution 
of the Laplace equation for $\psi_S$ in the solid and the pair of 
equations~\ref{eq:net_motion} for $n^{\mathrm{net}}$ and $\psi$ in the liquid, 
coupled under BCs Eq.~\ref{eq:net_BC}, is straightforward. The general 
solutions for the homogeneous equations of $\psi_S$ (Laplace, finite at the 
origin) and of $n^{\mathrm{net}}$ (Helmholtz, finite at infinity) are 
well-known:
\begin{equation}\label{eq:gen_sol1}
\psi_S(r,\theta)=\sum_l A_l~P_l(\cos\theta)~r^l,\,\,\,\,\,\,\,\,\,
n^{\mathrm{net}}(r,\theta)=\sum_l
B_l~P_l(\cos\theta)~k_l(\beta r),
\end{equation}
where $P_l$ are Legendre polynomials and $k_l$ are modified spherical Bessel 
functions of the second kind. The inhomogeneous Poisson equation for $\psi$ has 
an obvious special solution due to its structural similarity to the Helmholtz 
equation for $n^{\mathrm{net}}$: 
$\psi^{\mathrm{special}}=-(1/\beta^2)(e/\epsilon_0\epsilon'_w)n^{\mathrm{net}}$.
The general solution for the homogeneous Laplace equation, for which 
$\psi^{\mathrm{hom}}$ becomes the driving potential $-E_0 r\cos(\theta)$ for 
$r\gg a$, is simply $-E_0 r P_1(\cos\theta)+\sum C_l P_l(\cos\theta) r^{-l-1}$.  
Thus, the general solution for potential $\psi$ in the liquid is:
\begin{equation}\label{eq:gen_sol2}
	\psi(r,\theta)=-E_0 r P_1(\cos\theta)+\sum_l
	P_l(\cos\theta)\left(C_l~ 
	r^{-l-1}+\left(-\frac{e}{\epsilon_0\epsilon'_w}\frac{1}{\beta^2}\right)
	B_l~k_l(\beta r) \right).
\end{equation}
Combining Eq.~\ref{eq:gen_sol1} and Eq.~\ref{eq:gen_sol2} with BCs.  
Eq.~\ref{eq:net_BC}, where the normal current is simply the radial derivative 
against $r$, it is easy to see that for any angular eigenvalue $l\neq1$, 
because of the absence of external drive term $E_0$, the three BCs produce 
three homogeneous linear equations with a non-zero determinant and thus force 
$A_l=B_l=C_l=0$.  The remaining three $l=1$ coefficients can be easily solved 
through matching BCs, giving an exact solution for the potential and charge 
distribution in the liquid and the sphere.

It is crucial to note that the simple grouping in Eq.~\ref{eq:gen_sol2} by 
their angular eigenvalues $l$ depends crucially on that $P_l$ are \emph{shared 
}eigenfunctions for the angular part of axially-symmetric Laplace and Helmholtz 
equation. The same trick does \emph{not} work for even the slightly less 
symmetric geometry of a spheroid whose symmetry axis aligns with the direction 
of the driving field $\vec{E}_0$. For the spheroid geometry the coupled 
equations of the Laplace Equation for $\psi_S$ and Eq.~\ref{eq:net_motion} 
under BCs Eq.~\ref{eq:net_BC} are \emph{non-separable}.

The three terms of the potential $\psi$ in the liquid has straightforward 
physical interpretations. $-E_0 r \cos\theta$ is the driving field. The term 
with modified spherical Bessel function $k_l(\beta r)$ decays exponentially 
from the interface $\mathbf{\Sigma}$ over the scale $|1/\beta|\approx\lambda\ll 
a$. The term $C_1\cos\theta/r^2$ are precisely that of an induced dipole. If we 
define a polarization $P$ by the far field potential $\psi(|\vec{r}|\gg a)\to 
P\,a^3\,\vec{E_0}\cdot\vec{r}/(|\vec{r}|^3)-\vec{E}_0\cdot\vec{r}$, the exact 
solution above gives the following form of $P$:
\begin{equation}\label{eq:solution}
	P=1-\frac{3}{2+[\epsilon_w(\omega)/\epsilon_s(\omega)+g/i\omega\tau_D]^{-1}},\,\, 
	g=\left(a\beta+1+\frac{1}{a\beta+1}\right)^{-1},\,\,
	\epsilon_{w/s}(\omega)=
	\epsilon'_{w/s}+\frac{\sigma_{w/s}}{i\omega\epsilon_0},
\end{equation}
where the last equation is simply the standard form of the low-frequency 
complex dielectric constants for the liquid and the solid with the assumptions 
of frequency-independent $\epsilon',\sigma$.

This exact solution for $P$ can be simplified considerably under the low 
frequency regime in this paper. First, as noted below Eq.~\ref{eq:net_motion}, 
we focus on $\omega\tau_D\ll1$ so $\beta\approx1/\lambda$. As $\lambda$ is of 
the order 10nm even for dilute electrolyte solution, the radius of spheres 
$a\gg\lambda$ even for micron-sized microscopic particles, therefore 
$a\beta\approx a/\lambda\gg1$, $g\approx\lambda/a$ in Eq.~\ref{eq:solution} 
above. Furthermore, we can rewrite the ratio between complex dielectric 
constants as
\begin{equation}\label{eq:ratio}
	\frac{\epsilon_w(\omega)}{\epsilon_s(\omega)}=
	\frac{\sigma_w}{\sigma_s}\frac{1+i\omega\tau_D}{1+i\omega\tau_S}
	\approx\frac{\sigma_w}{\sigma_s},\,\,\,\,\,
	\omega\tau_D\ll1,\,\,\,\,\, \omega\tau_S=
	\omega\tau_D\frac{\epsilon_s}{\epsilon_{w}}\frac{\sigma_w}{\sigma_s}
	\ll1.
\end{equation}
Here the definition of $\tau_S$ for the solid is completely analogous to that 
of $\tau_D$ in Eq.~\ref{eq:net_motion}. As $\epsilon'_s/\epsilon'_w$ is 
generally between 0.01 and 0.1, as long as the solid is \emph{non-insulating}, 
the second condition above is not much more stringent than $\omega\tau_D\ll1$.

With these two approximations, the exact solution for $P$ reduces to a 
``Debye-like'' form:
\begin{equation}\label{eq:polarization}
	P=P_0+\frac{P_1}{1+i\omega\tau_C},\,\,\,\,
	P_0=1-\frac{3~\sigma_w/\sigma_s}{1+2~\sigma_w/\sigma_s},\,\,\,\, 
	P_1+P_0=-\frac{1}{2},\,\,\,\,
	\tau_C=\tau_D~\frac{a}{\lambda}
	\left(\frac{\sigma_w}{\sigma_s}+\frac{1}{2}\right).
\end{equation}
As we shall see below, the new time scale $\tau_C$, geometrically enhanced by a 
huge factor $a/\lambda$ profoundly alters the physics and gives rise to the 
dielectric enhancement.
\section{Dielectric Response for a Random Suspension of Spheres in an 
Electrolyte Liquid}\label{sec:suspension}
We now proceed to evaluate the linear response of a dilute suspension of 
spheres in Sec.~\ref{sec:single}, with a volume fraction $f\ll1$, in the 
electrolyte solution.  For this we employ the Maxwell~\cite{Maxwell} or the 
Maxwell-Garnett~\cite{MaxwellGarnett, Landauer, Batchelor} effective medium 
approximation. In this theory, the effective complex dielectric constant
$\epsilon_{\mathrm{eff}}(\omega)$ obeys the Clausius-Mossotti 
Relation~\cite{Landauer,Batchelor}:
\begin{equation}\label{eq:clausius}
fP=\frac{\epsilon_{\mathrm{eff}}/\epsilon_w(\omega)-1}{\epsilon_{\mathrm{eff}}/\epsilon_w(\omega)+2},\,\,\,\,
\frac{\epsilon_{\mathrm{eff}}}{\epsilon_w(\omega)}\approx1+3fP,\,\,\,\,
\frac{\epsilon_{\mathrm{eff}}}{\epsilon'_w}=\left(1+\frac{1}{i\omega\tau_D}\right)(1+3fP)\approx\frac{1}{i\omega\tau_D}(1+3fP).
\end{equation}
In the first step we use the fact that when $f\ll1$, $|fP|\ll1$, which is easy 
to verify with Eq.~\ref{eq:polarization}. The next step we simply use the fact 
that at low frequency $\omega\tau_D\ll1$, $\epsilon_w(\omega)$ is dominated by 
its imaginary parts from conduction.

Putting Eq.~\ref{eq:polarization} and Eq.~\ref{eq:clausius} together, we obtain 
the following Debye form of effective dielectric constant for a dilute $f\ll1$ 
suspension of non-insulating spheres under the two low-frequency conditions 
Eq.~\ref{eq:ratio}:
\begin{equation}\label{eq:effective}
	\frac{\epsilon_{\mathrm{eff}}}{\epsilon'_w}\approx
	\left(1-\frac{3}{2}f\right)\frac{1}{i\omega\tau_D}+
	\frac{9f}{4}\frac{a}{\lambda}\frac{1}{1+i\omega\tau_C}.
\end{equation}
The first term in $\epsilon_{\mathrm{eff}}$ is simply the divergent imaginary 
part of 
$\epsilon_w(\omega)=\epsilon'_w+\sigma_w/(i\omega\epsilon_0)=\epsilon'_w[1+1/(\omega\tau_D)]$
of the electrolyte solution, now slightly modified by the dilute suspension 
(the $3f/2$ term). This terms is purely imaginary, so 
Re$(\epsilon_{\mathrm{eff}}(\omega)$ is entirely determined by the second term.

That second term is rather more consequential. At low frequency, it produces a 
large, material-independent enhancement to the static dielectric constant:
\begin{equation}\label{eq:enhancement}
	\mathrm{Re}(\epsilon_{\mathrm{eff}}(\omega))=\left[\left(1-\frac{3}{2}f\right)
	+\frac{9}{4}\frac{a}{\lambda}f\,\right]\epsilon'_w,\,\,\,\,\,\,\,
	\omega\tau_C\ll1,\,\,\,or\,\,\,\,
	\omega\tau_D\ll\frac{\lambda}{a}
	\frac{1}{\sigma_w/\sigma_s+1/2}.
\end{equation}
Here we have included, for completeness's sake, insignificant contributions 
$(1-3f/2)$ from the terms neglected in the last step of Eq.~\ref{eq:clausius}.
The enhancement, scaling with $a/\lambda$ and independent of the material 
properties, such as $\epsilon_s$ and $\sigma_s$ of the solids, is determined 
solely by geometry. The condition for observing the enhancement is governed not 
by $\tau_D$, but by a new time scale dependent on both the geometry and the 
conductivity of the solid spheres.

We will discuss the physical origin of this time scale in more details below.  
Here we only note that, the larger is the geometric enhancement factor 
$a/\lambda$, the correspondingly lower one has to go to observe the 
enhancement.  Furthermore, if the solid is insulating $\sigma_s\to0$, the 
condition in Eq.~\ref{eq:enhancement} can never be satisfied and there is no 
dielectric enhancement.
\section{Scale-Invariant Maximum of Phase Shift}\label{sec:phase}
In general the dielectric measurements at low-frequencies are exceedingly
difficult\cite{Schwan}, precisely because the imaginary contributions from 
conduction dominates over the real part of the dielectric constant. The large 
enhancement predicted in the last section, however, should make this 
observation easier. To show this, let us compute the phase shift in complex 
dielectric constant, defined as 
$\tan\theta=\arg\epsilon_{\mathrm{eff}}(\omega)=\mathrm{Re}(\epsilon_{\mathrm{eff}})/\mathrm{Im}(\epsilon_{\mathrm{eff}})$, 
which is easily observable in simple impedance measurements.

For pure electrolyte solution without the solids, the phase shift is at low 
frequency is simply $\omega\tau_D$. As $\tau_D$ is $10^{-7}s$ even for a very 
dilute electrolyte solution, and $\omega\tau_D\ll1$, the phase shift is about 
0.1 miliradians for frequencies at kHz range. The dielectric enhancement of 
$(9f/4)(a/\lambda)$ boost this low frequency phase shifts by orders of 
magnitude with the introduction of an even dilute suspension of spheres, making 
it much more observable.

More quantitatively, let us rewrite Eq.~\ref{eq:effective} to this more 
convenient form:
\begin{eqnarray}\label{eq:phase-Debye}
	\frac{\epsilon_{\mathrm{eff}}}{\epsilon_w'}&\approx&\frac{a}{\lambda}\left(\sigma_w/\sigma_s+\frac{1}{2}\right)
	\left[\left(1-\frac{3f}{2}\right)\frac{1}{i\omega\tau_C}
	+\frac{9f}{4\sigma_w/\sigma_s+2}\,\frac{1}{1+i\omega\tau_C}\right]\nonumber\\
	&\approx&\frac{a}{\lambda}\left(\sigma_w/\sigma_s+\frac{1}{2}\right)
\left[\frac{1}{i\omega\tau_C}
	+\frac{9f}{4\sigma_w/\sigma_s+2}\,\frac{1}{1+(\omega\tau_C)^2}\right]
\end{eqnarray}
The second step relies on the fact that $f\ll1$, so both the $-3f/2 
(1/i\omega\tau_D)$ contribution and the imaginary part of the Debye term are 
small compared with $1/(i\omega\tau_D)$. Now it is trivial to see that the 
phase shift and its maximum are:
\begin{equation}\label{eq:phase}
	\tan\,\theta(\omega)=\frac{9f}{4\sigma_w/\sigma_s+2}\,\frac{\omega\tau_C}{1+(\omega\tau_C)^2},\,\,\,\,\,\,\,\,
	\tan\,\theta(\omega_{\mathrm{max}})=\frac{9f}{4}\frac{1}{2\sigma_w/\sigma_s+1}.
\end{equation}
The maximum of phase shift is observed at frequency 
$\omega_{\mathrm{max}}\tau_C=1$, or spelling out the geometric and material 
dependence explicitly:
\begin{equation}\label{eq:phase-max}
	\omega_{\mathrm{max}}\tau_D=\frac{2\lambda}{a}\frac{1}{1+2\sigma_w/\sigma_s},\,\,\,or\,\,\,
	\omega_{\mathrm{max}}=\frac{2D}{\lambda 
a}\frac{1}{1+2\sigma_w/\sigma_s}.
\end{equation}
Two features stands out from these results. The maximum phase shift is 
\emph{scale-invariant}.  Indeed, given that even strong electrolyte solutions 
like the sea water and the human blood has conductivity $\sigma_w$ of the order 
$S/m$, and even moderately doped semiconductor has $\sigma_s$ orders of 
magnitudes higher, it is very often safe to assume $\sigma_s\gg\sigma_w$ for 
non-insulating solids.  In such a case, the maximum phase shift depends 
\emph{solely} on volume faction $f$ and is very significant: even for a small 
volume faction $f=0.05$, $\theta$ reaches over 100 miliradian. Dielectric 
enhancement from even a dilute suspension of non-insulating spheres boosts the 
minuscule phase shift of an electrolyte solution to an easily observable level. 

The second feature is that the frequency $\omega_{\mathrm{max}}$ at which the 
maximum above is observed is inversely proportional to the size of the spheres.  
This is different from the scaling of the maximum frequency of phase shift 
observed in clay particles 
$\omega_{\mathrm{max}}=D/a^2$~\cite{Dukhin,Fixman,Chew,Hinch}.  The mechanism 
for dielectric enhancement in that case, the large $\zeta$-potentials and 
surface charges, is entirely different from ours.

\section{Effective Boundary Conditions and the Physical 
Picture}\label{sec:physics}
From the Sec.~\ref{sec:single} and Sec.~\ref{sec:suspension} above, we see that 
the origin of the dielectric enhancement is the small but non-vanishing 
imaginary part in the dielectric response of the sphere $P$ that \emph{persists 
at low frequency }$\omega\tau_D\ll1$. That imaginary part, the 
$g/(i\omega\tau_D)$ in Eq.~\ref{eq:solution} which gives rise to the 
$i\omega\tau_C$ in Eq.~\ref{eq:polarization}, form a \emph{cross term} that, 
when multiplied by the \emph{diverging } imaginary conduction contribution from 
the electrolyte solution $1/(i\omega\tau_D)$ in Eq.~\ref{eq:clausius}, produces 
the large \emph{real} dielectric constant at low frequency. As we show in 
Appendix I, without that term, at $\omega\tau_D\ll1$ and $\omega\tau_S\ll1$ 
(\emph{c.f.} Eq.~\ref{eq:ratio}) $P$ is purely real and there is no 
enhancement. So a physical picture of the enhancement need to account for this 
imaginary part.

From Eq.~\ref{eq:gen_sol1} and Eq.~\ref{eq:gen_sol2}, it is easy to observe 
that, in the liquid, the net charge $n^{\mathrm{net}}$ decays exponentially 
over the distance of order $\lambda$, and the potential $\psi$, aside from the 
long range parts due to the external drive and the induced dipole, also does 
so. It is then natural to separate out the physics within these thin, charged 
layers within a few $\lambda$ from the solid-liquid interfaces, which we call 
the ``\underline{E}xteranlly \underline{I}nduced \underline{D}ipole 
\underline{L}ayers''(EIDL), from the \emph{charge-neutral} homogeneous bulk 
liquid outside it.  The potential $\psi_L$ in the neutral bulk liquid, just 
like the potential within the solid $\psi_S$, follows the much simpler Laplace 
equation, and is thus a harmonic function.

By analyzing the charge dynamics and potential within the EIDL, we can derive 
effective boundary conditions (BCs) connecting harmonic potentials $\psi_S$ and 
$\psi_L$,:
\begin{equation}\label{eq:BC_eff}
	\epsilon_{s}(\omega) E^{\bot}_{S}=\epsilon_w(\omega) E^{\bot}_{L}
	\,\,\,\,\,\,\mathbf{[a]},\,\,\,\,\,
	\psi_{S}=\psi_{L}+\frac{1}{i\omega\tau_D\beta } E^{\bot}_{L}.
\,\,\,\,\mathbf{[b]},
\end{equation}
where $\epsilon_{w/s}(\omega)$ are defined in Eq.~\ref{eq:solution}.

While detailed derivation of these BCs can be found in another 
publication~\cite{Jiang}, the basic idea is straightforward. It is easy to see 
from Eq.~\ref{eq:gen_sol1} and Eq.~\ref{eq:gen_sol2} that the spatial variation 
within the EIDL is much more rapid in the normal direction, where both 
$n^{\mathrm{net}}$ and $\psi$ vary on the scale of $\lambda$, than in the 
tangential direction, where they vary on the scale of $a$. The governing 
equations Eq.~\ref{eq:net_motion} therefore reduce into 1D problems in the 
normal direction $\hat{u}$, and each spatial derivative simply contributes a 
factor of $-\beta$, just as each time derivative contributes a factor of 
$i\omega$ under a harmonic drive.

Furthermore, the normal net current $j^{\mathrm{net}}=j^+-j^-$ consists of a 
constant contribution from outside the EIDL $j^{\mathrm{out}}=\sigma_w 
E^{\bot}_L$ ($E^{\bot}_L$ is the normal field in the charge neutral liquid 
\emph{just outside} the EIDL) and a current $j^{\mathrm{var}}$ from within the 
EIDL that varies spatially as $\exp(-\beta z)$. Charge conservation $\partial 
n^{\mathrm{net}}/\partial t+\vec{\nabla}j=0$ then dictates $\beta 
j^{\mathrm{var}}=i\omega n^{\mathrm{net}}$ throughout the EIDL. Finally, the 
critical boundary condition that the normal current vanishes at the 
liquid-solid interface $j=j^{\mathrm{var}}+j^{\mathrm{out}}=0$ (\emph{c.f.} 
Eq.~\ref{eq:bc}a) implies that the net particle density at that interface $n_0$ 
satisfies:
\begin{equation}\label{eq:n_0}
j^{\mathrm{out}}=\sigma_w E^{\bot}_L,\,\,\,\,\,\,
\beta j^{\mathrm{var}}=i\omega n^{\mathrm{net}},\,\,\,\,\,\,
j=j^{\mathrm{var}}+j^{\mathrm{out}}=0,\,\,\,\,\,\,
n_0=-\frac{E^{\bot}\beta\sigma_w}{i\omega e}
\end{equation}
Since within the EIDL $n(z)=n_0\exp(-\beta z)$, with one spatial integration of 
the Poisson equation one can find how the normal field $E^{\bot}$ varies across 
the EIDL, and with another integration one can obtain the potential drop across 
it.

The physical interpretation of the effective BCs Eq.~\ref{eq:BC_eff} is 
straightforward. Recall that in Eq.~\ref{eq:bc}~[{\bf c}], the conduction 
current is missing at the boundary between the solid-liquid interfaces.  
Eq.~\ref{eq:BC_eff}~[{\bf a}] shows that, after taking into account of the 
charge dynamics within the EIDL, the ordinary BC for harmonic potentials on 
conductive boundaries is restored between the harmonic potentials $\psi_S$ and 
$\psi_L$ on either side of the EIDL.

The second BC Eq.~\ref{eq:BC_eff}~[{\bf b}] is more significant. The 
discontinuity between $\psi_L$ and $\psi_S$ comes from the dipole moment 
produced by a large, inhomogeneous charge build-up within the EIDL. This dipole 
moment is out of phase (purely imaginary) from the driving field, because the 
driving field is proportional to the $j^{\mathrm{out}}$ in the EIDL 
(Eq.~\ref{eq:n_0}), whereas the charge dynamics is always within the EIDL is 
\emph{always out of phase }
from $j^{\mathrm{var}}$ (\emph{ibid.}), and the two currents are linked by the 
perfectly polarizable boundary condition $j=0$. As we will see below, it is 
precisely this out of phase dynamics that generates the imaginary part of $P$ 
in Eq.~\ref{eq:polarization} and thus produces the dielectric enhancement.
Furthermore, as the driving frequency $\omega$ decreases, the charge dynamics 
slows and this forces a larger charge build up $n_0$ within one period 
$T=2\pi/\omega$ in the EIDL (Eq.~\ref{eq:n_0}). Thus the induced dipole of the 
EIDL and  the dielectric enhancement are inherently low-frequency phenomena.

Finally, the special symmetry of the sphere geometry permits a particularly 
simple solution from the effective BCs Eq.~\ref{eq:BC_eff}. As noted in 
Sec.~\ref{sec:single}, due to the symmetry of the driving field, only terms 
with $P_1=\cos\theta$ angular dependence survive in all potentials and 
densities.  In particular, the electric field $E_S$ within the solid sphere is 
uniform and parallel to the external drive. This means that, on the sphere 
surface, the normal electric field at polar coordinate $(\theta,\phi)$ is 
simply $E_S\cos\theta$. Setting $\psi_S=0$ at the origin, the potential at 
$(\theta,\phi)$ is $-Ea\cos\theta$.  Thus, for a sphere in a uniform driving 
field, $\psi_S/E^{\bot}_S=-a$, \emph{throughout } the solid-liquid boundary 
$\mathbf{\Sigma}$. This allows us to convert the dipole term in 
Eq.~\ref{eq:BC_eff}~[{\bf b}] back to $\psi_S$ via Eq.~\ref{eq:BC_eff}~[{\bf 
a}].  By rescaling rescaled internal potential and field with a proper factor 
to $\psi_S'$ and $E_S'$, we can transform the boundary conditions 
Eq.~\ref{eq:BC_eff} into a ``conventional'' boundary value problem, without the 
dipole layer between two homogeneous medium, that leaves the potential outside 
$\psi_L$ \emph{unchanged},  only with a new relative permittivity:
\begin{equation}\label{eq:BC_rescale}
	\epsilon'(\omega)E'^{\bot}_S=\epsilon_w(\omega)E^{\bot}_L,\,\,\,
	\psi'_L=\psi_L,\,\,\,
	\frac{\epsilon_w(\omega)}{\epsilon'(\omega)}=
	\frac{\epsilon_w(\omega)}{\epsilon_s(\omega)}+\frac{\lambda}{a}\frac{1}{i\omega\tau_D}.
\end{equation}
From elementary electrostatics we know that for a sphere with dielectric 
constant $\epsilon'$ immersed in an uniform medium with $\epsilon_w$, the 
induced polarization, as defined above Eq.~\ref{eq:solution} 
$P=1-3/(\epsilon'/\epsilon_w+2)$. With the peculiar relatively permittivity in 
Eq.~\ref{eq:BC_rescale}, this elementary result immediately leads to the exact 
solution Eq.~\ref{eq:solution}. The sole difference is the reduction of $g$ in 
Eq.~\ref{eq:solution} to $\lambda/a$, which is valid when $\omega\tau_D\ll1$ 
and $\lambda\ll a$. These are precisely the conditions for effective BCs 
Eq.~\ref{eq:BC_eff} in the EIDL approximation~\cite{Jiang}. This solution also 
makes it clear that the origin of the imaginary term in $P$ at low frequency
$\omega\tau_D\ll1$ is due solely to the imaginary dipole layer term in 
Eq.~\ref{eq:BC_eff}~[{\bf b}].

\section{Application to Experiments}\label{sec:experiment}

There are numerous experiments that show dielectric enhancement in metallic 
particles. IP is an important tool in  prospecting for copper. Porphyry copper 
deposits, for example,   contain  disseminated conducting  chalcocite 
($Cu_2S$), chalcopyrite ($CuFeS_2$), and pyrite ($FeS_2$) embedded in a porous 
matrix that is predominantly feldspar, quartz, and mica. The conducting grains, 
in most types of mineral deposits, are in contact with brine that fills the 
pore space, and, produce huge IP signals in the field measurements.  Hence 
there are quite a large number of laboratory data on these materials.  
Wong\cite{Wong} cites data by numerous authors; see also the references cited 
in \cite{Nelson, Collett} as well as those cited by Seigel et al.\cite{Seigel} 
in their review of the history of IP. In many ways these experiments are 
uncontrolled as many important parameters such as pH, reactions effects, if 
any,   the $\zeta$-potential were not measured separately. More relevant for 
our model of mono-sized spherical grains, The shapes and homogeneity of the 
particles are not guaranteed.  Nelson and Van Voohris\cite{Nelson} emphasize 
that the deficiency in their data lies in the fact that the cation exchange 
capacity (CEC), that is directly related to the $\zeta$-potential, were not 
measured. 

Figure 14. in Ravel  \cite{Nelson} shows a typical phase-shift and in-phase 
conductivity for Pyrite content of 1 percent  in weight (10 mg/g) in sand or in 
agar gel. It has  features that are not unlike the figure below,  
Fig(\ref{fig:experiment}). As mentioned above, the phase shift maximum varies 
inversely with the particle size\cite{Wong, Collett}, for the smaller 
particles,  and data agrees with Eq. (\ref{eq:phase-max}). It can be surmised 
that these may be ideally polarizable samples.  The data in the geophysical 
literature show dielectric enhancement in many metallic and semiconducting 
particles and  the height in the phase shift maximum depends linearly on the 
volume concentratio\cite{Wong, Nelson}.  Preliminary experiments with 
conductive particles  that have putatively zero zeta  potential such as iron 
and carbonaceous particles(``coke-breeze") \cite{Klienhammes} show the behavior 
predicted here--$\tan{\theta}$   depends linearly on the volume fraction $f$. 
\begin{figure}
\begin{center}
{\includegraphics[width=1.0\textwidth]{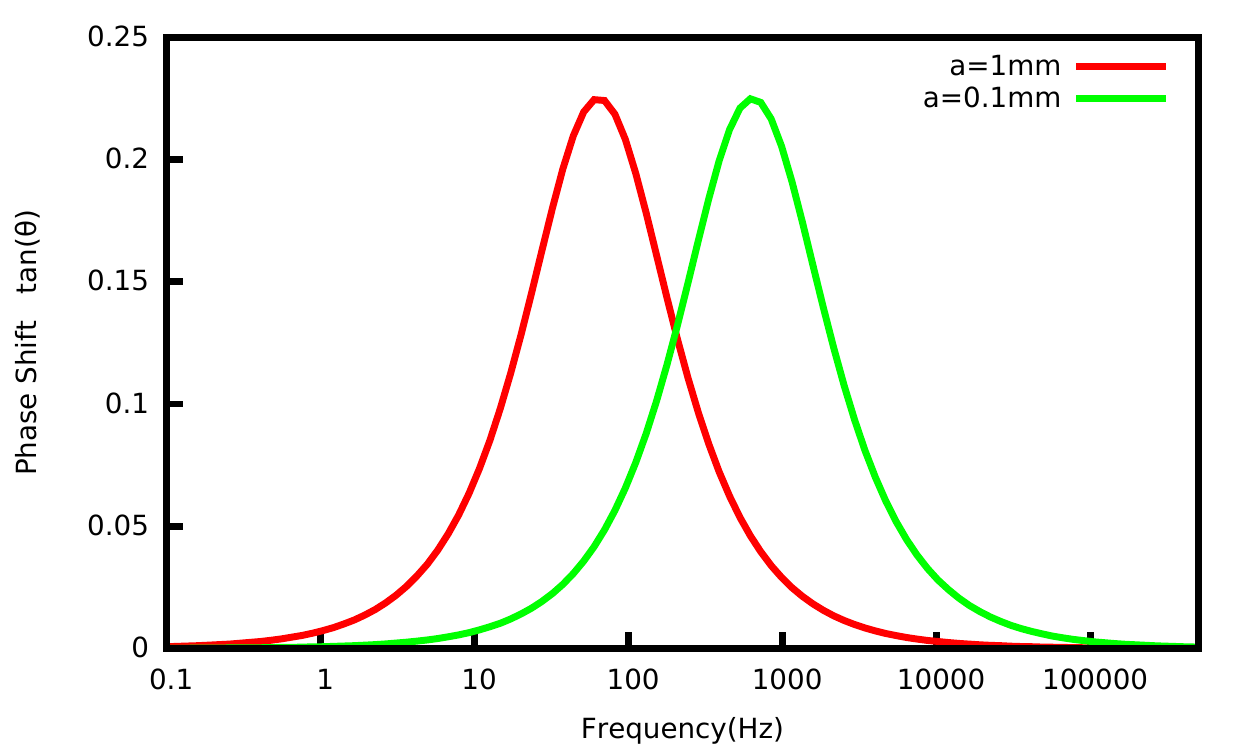}}
\end{center}
 \caption[]{ A plot of $\tan{\theta}$ in Eq. (\ref{eq:phase}), for volume 
 fraction $f=0.1$ and electrolyte concentration $10^{-3}M$, as a function of 
 the frequency $\omega$. Spheres of radius $a=0.1mm,1mm$ are plotted. It was 
 assumed that the material conductivity is much greater than that of water.  
 While the position of the maximum moves to higher frequencies, 
 $\omega_{\mathrm{max}}=2D/(\lambda a$), for a smaller  size, the height 
 remains the same as seen in experiment by \cite{Klienhammes}.  Also a linear 
 dependence on volume fraction $f$ is seen in the experimental data 
 \cite{Nelson, Collett, Seigel, Klienhammes}.}
 \label{fig:experiment}
\end{figure}

The maximum of $9f/4$ is easily observable by a commercial instrument, which 
often have $0.1$-milirad sensitivity, even for a very small volume fraction 
$f$. When designing contrast agent, one is often restricted by various 
practicality, like the frequency used by the apparatus or the in-situ 
conductivity. The above two equations Eqs. \emph{i.e.}, Eq. (\ref{eq:phase}) 
and Eq.( \ref{eq:phase-max}) will serve as a guide in designing the materials 
and their sizes. 

The dielectric spectroscopy  and the dielectric enhancement are  important 
tools in biology \cite{Cole, Schwan, Blood}. Many complex macromolecules have 
fixed charges \cite{Cole} and have Guoy-Chapman like double layer and the 
dielectric enhancement can be likened to those  of clay \cite{Dukhin, Fixman, 
Chew, Hinch}.   The dielectric spectroscopy has been used widely in other 
biological systems like blood. However   blood has strong ionic conductivity.  
The dielectric constant generally increases with lowering frequency from the 
dielectric of the host and ending with a plateau with the enhanced value. Blood 
shows rather complicated dielectric response\cite{Blood} with two plateaus, as 
the frequency is lowered. It has been suggested that the  plateau lower 
frequency enhancement is perhaps due to and ``blocking" effect. This requires 
more careful examination as the membranes are not impermeable to ions. The  
above analysis needs to be be extended to incorporate both finite 
$\zeta$-potential and finite permeability of the membrane.

Finally, we would like to draw attention of the reader to Ref.\cite{Jiang} 
where we show that the dielectric enhancement is nearly universal for ideally 
polarizable interfaces \emph{i.e.} homogeneous particles of arbitrary shape 
suspended in an electrolyte solution. Furthermore, in this paper, to emphasize 
the physics and simplify notations, the solution contains only one species
each of cations and anions, with equal and opposite charges   and equal
diffusion diffusion coefficients of the cations and anions are $D_+,D_-$. We 
show in \cite{Jiang} that asymmetric ions with distinct charges and diffusion 
coefficients have the same physics. We also show in \cite{Jiang} that  the 
Electrophoretic Flow is insignificance in the present problem of dielectric 
enhancement.
 
\section{Acknowldgements}
At the early stages of this work, one of us (Sen) was partially supported at 
the University of North Carolina, Chapel Hill,  by the Advanced Energy 
Consortium: http://www.beg.utexas.edu/aec/ whose  member companies include BP 
America Inc., Total, Shell,  Petrobras, Statoil, Repsol,  and Schlumberger. Sen 
is   grateful to  Alfred Kleinhammes   for stimulating discussions and 
collaborations on experiments. 

\bigskip
\bigskip

\textbf{ Appendix I: Absence of Dielectric Enhancement without Ionic Effects} 

\bigskip
\bigskip

In the text-book \cite{Maxwell} examples of potential induced by a sphere with 
zero $\zeta$-potential, the conductivity in the host is driven by the 
electrical potential gradient times $\sigma_{w}$ and there are no  currents  
from the gradients in the carrier densities. The potentials both inside and  
outside the sphere are  governed by the Laplace's equation, i. e.  the charge 
imbalance is zero in the Poisson's equation. In this case although there is an 
induced surface charge density on the surface of the sphere--there is no EIDL.
We briefly recollect that without a double layer like EIDL or a Guoy-Chapman layer,  there is no enhancement.  For a sphere of  dielectric $\epsilon_{in}$,  suspended in a continuum $\epsilon_{w}$
\begin{equation}
\label{ordinaryP }
P    =   \frac{\epsilon_{in} -\epsilon_w}{\epsilon + 2 \epsilon_w },
\end{equation}
 as given in \cite{Maxwell}.

For inclusions with non-zero conductivity $\epsilon_{in} 
=\epsilon^{\prime}_{in}+ \sigma _{in}/i \omega  \epsilon _0  $, we have,  at 
low frequencies, using the method outline in the text, 
Eq.(\ref{eq:clausius})
\begin{equation}
\label{ MGTcross}
\epsilon_{\mathrm{eff}} \rightarrow 3 f \frac{ \sigma_w}{\omega\epsilon_0 }   
\frac{P^{\prime\prime}}{a^3} =\frac{9 f \sigma _w \left(\sigma _{in} 
\epsilon^{\prime} _w-\epsilon^{\prime} _{in} \sigma _w\right)}{\left(2 \sigma 
_{in}+\sigma _w\right){}^2}=9 f  \epsilon^{\prime} _w \frac{ \left(\frac{\sigma 
_{in}}{\sigma _w} -\frac{\epsilon^{\prime} _{in}}{\epsilon^{\prime} 
_w}\right)}{\left(2 \frac{\sigma _{in}}{\sigma _w}+1\right){}^2}\end{equation}
The above equation shows  that for any choice of the parameters  there is \emph{no} significant enhancement.

For  insulating particles (without surface charges)  embedded in conducting 
fluid at low frequencies  $P=   ({\epsilon_{in}-\epsilon_w})/({\epsilon_i+ 2 
\epsilon_w})  \rightarrow -1/2  $ and we recover Maxwell's celebrated result 
that inserting insulating particles reduces the overall low frequency 
dielectric constant
\begin{equation}
\label{effective5}
\epsilon_{eff}= \epsilon_w (1- \frac{3}{2} f)  
\end{equation}

\bigskip
\bigskip

\textbf{ Appendix II: Planar electrodes } 

\bigskip
\bigskip
We now briefly contrast our results with those that are known for planar 
electrodes where the   enhancement  mars most of the two-electrode dielectric 
measurements (at low frequency) and makes a four probe measurement essential.  
The results bear some similarity to the results derived in the text.

The planar electrodes with ideally polarizable interface also shows a 
dielectric enhancement, not unlike its more well  interfacial redox reactions 
(Faradaic effect) dominated counterpart known as Warburg 
impedance\cite{Kohlrauch, Warburg}. The subject of ideally polarized planar 
electrodes, by itself, is an enormously important problem in electrochemistry 
\cite{Bard}. We refer to  the skillful and succinct review by 
Hollingsworth\cite{Schwan, McDonald, Barbero, Hollingsworth} and references 
therein.    

For two  planar ideally polarized electrodes separated by a distance $d$,  with 
$
\tau= \tau_D ( d  / 2 \lambda)= d \lambda /(2D)=\sqrt{  \tau_D \tau_L}\;\; ; \tau_L=\frac{(d/2)^2}{D}$.
the effective dielectric constant has a canonical Debye form  
\begin{equation}
\label{Debye2}
\epsilon^{\prime}_{eff}   =    \epsilon^{\prime}_w \epsilon_0  + \frac{\Delta \epsilon^{\prime} \, \epsilon_0 }   {1+ i \omega   \tau } ; \,\,\,\Delta \epsilon^{\prime}  = \epsilon^{\prime}_w  \frac{d}{ 2 \lambda} ; \,\,\, \tau=  \frac{ d \lambda}{2D},  
\end{equation}  
 
Thus, at low frequencies, $\epsilon'_{eff} (\omega)$, the real part of 
$\epsilon_{eff}$ exceeds the water dielectric constant $\epsilon^{\prime}_w$  
by the factor $d/(2 \lambda)$. With  $\tau_D \sim 10^{-6} $sec  and  for mm 
size separation $d/\lambda \sim 10^6$, an  enhancement of $\epsilon'_{eff}  
\sim 10^6 \epsilon_w$ can happen  for frequencies such that $ 
(d/\lambda)(\omega \tau_D) <1 $ .i.e.   $\omega \lesssim 1 Hz $. 

The results for the planar electrodes are not unlike the results for the 
sphere: the time constant  and enhancements  are given by the correspondence $d 
\leftrightarrow a$, as can be expected from a dimensional analysis\cite{Jiang}.


\end{document}